# Anisotropy of the upper critical fields and the paramagnetic Meissner effect in $La_{1.85}Sr_{0.15}CuO_4$ single Crystals

I. Felner[*] and M.I. Tsindlekht
Racah Institute of Physics, The Hebrew University, Jerusalem, 91904, Israel

G. Drachuck and A. Keren,
Department of Physics, Technion - Israel Institute of Technology, Haifa, 32000, Israel.

Optimally-doped $La_{1.85}Sr_{0.15}CuO_4$ single crystals have been investigated by dc and ac magnetic measurements. These crystals have rectangular needle-like shapes with the long needle axis parallel to the crystallographic *c* axis (*c*-crystal) or parallel to the basal planes (*a*-crystal). In both crystals, the temperature dependence of the upper critical fields ($H_{C2}$) and the surface critical field ($H_{C3}$) were measured. The H-T phase diagram is presented. Close to $T_C$ =35 K, for the *c*-crystal, $\gamma^c = H^c_{C3}/H^c_{C2} = 1.80(2)$, whereas for the *a*-crystal the $\gamma^a = H^a_{C3}/H^a_{C2} = 4.0(2)$ obtained, is much higher than the theoretical value 1.69. At low applied dc fields, positive field-cooled branches known as the "paramagnetic Meissner effect" (PME) are observed, their magnitude is inversely proportional to H. The anisotropic PME is observed in both *a*- and *c*-crystals, only when the applied field is along the basal planes. It is speculated that the high $\gamma^a$ and the PME are connected to each other

PASC numbers: 74.25.Dw, 74.25.Ha, 74.25.Op, 74.90.+n

**Introduction:**

Bulk superconductivity (SC) in type-II superconductors appears when the applied magnetic field (H) is lower than the upper critical field $H_{C2}$. For optimally-doped $La_{1.85}Sr_{0.15}CuO_4$ (LSCO, $T_C \sim$ 35K) and $YBa_2Cu_3O_{7-\delta}$ (YBCO, $T_C \sim$ 92K) single crystals, the temperature dependence of $H_{C2}$ has been extensively studied only close to $T_C$, but their low temperature behaviour has not been fully understood. This is mainly due to the fact that at low temperatures, their normal state can be accessed only by magnetic fields higher than 50 T. Ultra high magnetic fields can be generated only for a short duration (less than 10ms), therefore conventional resistivity measurements are extremely difficult. The first low temperature study on YBCO up to 250 T for a magnetic applied field (H) parallel to the $CuO_2$ planes has been performed only very recently. [1] The constructed field-temperature phase diagram for YBCO, over a wide temperature range yields at 5 K $H_{C2}$ = 240 T (see [1] and in references therein). In contrast to YBCO, very few studies have been reported on $H_{C2}$(T) of LSCO single crystals and it was found that $H_{C2}$ is anisotropic. [2] Near $T_C$, the temperature dependence of $H_{C2}$ for H ∥ $CuO_2$ ($H^a_{C2}$) is higher than for H ∥ c ($H^c_{C2}$).[2] The estimated $H^a_{C2}$(0) and $H^c_{C2}$(0) values are 28 and 5.18 T respectively, thus the anisotropy ratio is 5.40.[3]

It is well accepted, that for H applied parallel to the sample surface, the transition to the SC state takes place at H<$H_{C3}$. For a single band superconductor, the theoretical ratio $\gamma = H_{C3}/H_{C2}$ predicted by Saint-James and De Gennes in 1963 is: ≈1.69. [4] That means that the nucleation field of a thin SC sheet at the surface, with a thickness of the order of the Ginzburg-Landau coherence length, is 70% higher than nucleation field of the bulk. Experiments on the conventional SC have confirmed that $\gamma$ is ~1.81. [5] It turns out that $\gamma$ is temperature and boundary conditions dependent. [6-7] So far, to our best knowledge, the surface superconducting state (SSS) in $La_{1.85}Sr_{0.15}CuO_4$ has not yet been studied. On the other hand, for the layered $K_{0.73}Fe_{1.68}Se_2$ single crystal (for H∥ab), the temperature dependence of $H_{C3}$ and $H_{C2}$, yields, $\gamma \sim 4.4$, [8] a value which is much larger than the predicted $\gamma = 1.69$ discussed above.

The Meissner effect, the expulsion of magnetic flux when a superconductor is placed in a magnetic field and cooled down through its $T_C$ (the field-cooled (FC) process) is arguably the finger print of the SC state. Surprisingly, several recent experiments have shown that at low H only, some SC materials may attract the magnetic field. It means that positive magnetization signals appear via the FC procedure. This is the so-called paramagnetic Meissner effect (PME). In YBCO single crystal the PME is small and was observed only when H was directed along the c-axis. [9] On the other hand in powder Bi-2212, the positive magnetic susceptibility reaches 60% of the complete diamagnetic Meissner effect.[10] The scarce experimental evidence for PME makes it difficult to identify the origin of this enigmatic phenomenon, although a large number of possible explanations have been advanced. [11-12] Although it is possible that the proposed mechanisms do play a role in high-$T_C$ superconductors, more recent observations of PME in Nb [13] clearly indicate the existence of another, less-exotic mechanism. [14] The limited choice of assumptions in this case makes the origin of PME more mysterious. The theory for conventional SC employs, that the PME originates from flux capture (caused by in-homogeneities) inside a SC sample and its consequent compression with decreasing temperature. [14] Alternatively, PME could also be an intrinsic property of any finite-size superconductor due to the presence of sample boundaries. The PME is explained by the giant vortex state with fixed orbital moment. Compressing of the magnetic flux trapped by this vortex state can lead to the PME. [15] A more recent theoretical model which is distinct from the previous ones, assumes that the PME is caused by impurities (such as oxygen vacancies) and the localized moments of the two-level system found in HTSC. This two level system is created by the surplus oxygen atom and neighbouring oxygen vacancy, united by a common singlet electron pair and their moment produces the PME effect. [16]

Optimally-doped $La_{1.85}Sr_{0.15}CuO_4$ single crystals have been grown in an image furnace. They were cut to rectangular needle-like shapes with the long needle axis parallel to the crystallographic *c*- axis or parallel to the *ab* basal planes, assigned as "*c*" and "*a*" crystals respectively, as depicted in Fig.1(inset). The resistivity and the low field susceptibility measurements of both crystals, clearly indicate anisotropy in the temperature at which the magnetization is detectable, and also anisotropy in the temperature at which zero resistivity appears.[17]. The major findings reported here are as follows. (i) The $H_{C2}(T)$ and $H_{C3}(T)$ data for the same "*a*" and "*c*" crystals, deduced by dc and ac magnetic fields respectively, for H parallel to the long dimension of the two crystals have been measured. It is shown, that near $T_C = 35$ K, for the *c*-crystal $\gamma^c = H^c_{C3}/H^c_{C2}$ is 1.8(2), a value which in fair agreement with the theoretical 1.69 ratio. [4] On the other hand, for the a-crystal the high $\gamma^a = H^a_{C3}/H^a_{C2} = 4.0(2)$ obtained, is very similar to that observed in the layered $K_{0.73}Fe_{1.68}Se_2$ measured in the same geometry. [8] (ii) We demonstrate anisotropic PME in $La_{1.85}Sr_{0.15}CuO_4$, which is observed in both *a*- and *c*- crystals, only when H is along the basal planes (H||*ab*). On the other hand, for H||*c* no PME is detected and the usual negative Meissner state is observed. It is speculated that these two new phenomena, namely (i) the high $\gamma^a$ ratio and (ii) the PME for H||*ab* only, are inter-connected to each other and have basically the same origin.

**Experimental Details**
A relatively large optimally doped $La_{1.85}Sr_{0.15}CuO_4$ single crystal has been grown in an image furnace. The crystal was oriented with an x-ray Laue camera, and then cut to rectangular needle-like shapes with the long needle axis parallel to the crystallographic *c*- axis (c-crystal) or parallel to the *ab* basal planes (a-crystal). [17] The dimensions are: 1.7*2.0*3.2 mm$^3$ for the c-crystal and 1.8*2.3*8.5 for the a-crystal. The composition, as well as the homogeneity of the crystals, were studied by various methods as described in Ref. 17. The dc magnetic study was done in a commercial MPMS5 Quantum Design superconducting quantum interface device (SQUID) magnetometer. The real ($\chi'$) and imaginary ($\chi''$) ac susceptibilities were measured with a home-made pickup coil method at amplitude of $h_0$=0.05 at various frequencies up to 1465 Hz.. The crystals were inserted into one of a balanced pair of coils, and the unbalanced signal was measured by a lock-in amplifier. The setup was adapted to the SQUID magnetometer as described in details in Ref. 18.

**Experimental Results**

Real ($\chi'$) and imaginary ($\chi''$) ac susceptibility measurements have been performed on both *a*- and *c*- crystals at several frequencies under various applied dc fields. All the ac susceptibility data presented here were collected when the dc field was applied parallel to the crystals' long needle axis. Both $\chi'$ and $\chi''$ signals are not affected by the frequency. Thus, the plots obtained for both crystals at 293 and 1465 Hz were identical. Fig. 1 shows the $\chi'$ and $\chi''$ curves measured at 1465 Hz at dc zero field. The onset of $\chi'$ signals are at $T_C$= 34.8(1) and 35.4(1) K for *a*- and *c*- crystals respectively, in good agreement with $T_C$=35 K, reported in Ref. 17. Generally speaking, the ac plots in Fig. 1 are in fair agreement with the resistivity and low field dc data reported in Ref. 17. The relatively broad transitions may indicate a small spread in stoichiometry along the crystals. This assumption is consistent with the double peaks observed in the $\chi''$ curve at 31.6 and 33.6 K for *a*- crystal and at 32.2 and 34.4 K for the *c*- crystal. Alternatively, the two steps in both $\chi'$ and $\chi''$, may be attributed to a different $T_C$ of the bulk and the surface, as discussed later. [6]

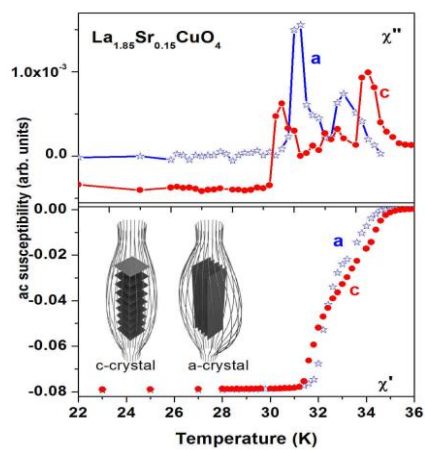

**Fig. 1**. Temperature dependencies of real ($\chi'$) and imaginary ($\chi''$) components of the ac susceptibility of La$_{1.85}$Sr$_{0.15}$CuO$_4$ for *a*- and *c*- single crystals measured at 1465 Hz at H=0. For the sake of clarity, the two ($\chi''$) curves were shifted from each other. The inset shows the schematic orientation of the *c*- and *a*- crystals.

**(i) Determination of $H_{C2}$(T)**

The criterion for determining the upper critical field $H_{C2}$(T) requires consistency, and not one method is entirely unambiguous. Here, $H_{C2}^a$(T) and $H_{C2}^c$(T) were determined by using two complementary methods. Isothermal field dependence of the dc magnetization M(H) has been measured at various temperatures. Two typical hysteresis loops measured at 28 and 30 K are presented in Fig. 2. $H_{C2}$ (indicated by the vertical arrow) was defined as the field at which the two ascending and descending curves merge. Alternatively, the temperature dependence of the magnetization M(T) under various applied fields has been measured and $H_{C2}(T)$ was determined at the crossing point as shown in Fig. 2 (inset). The two methods yield practically the same $H_{C2}^{c}(T)$ values, as presented in Fig.3.

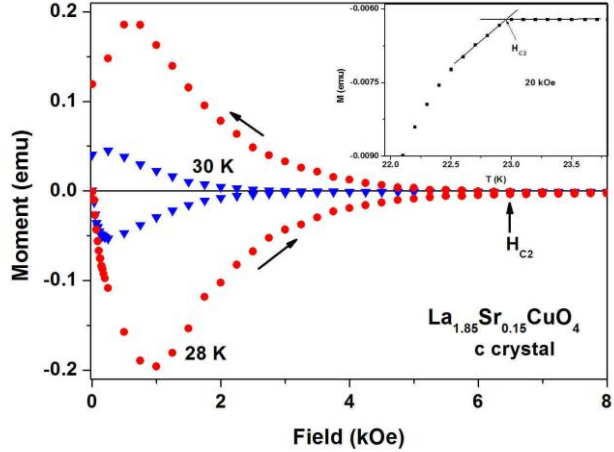

**Fig. 2**. Two isothermal dc magnetic hysteresis loops measured at 28 and 30 K for the *c*-crystal and the M(T) plot measured at 20 kOe (inset), from which $H_{C2}^{c}(T)$ is deduced.

A rough estimation of $H_{C2}(0)$ can be achieved by using the well-known Werthamar-Helfand-Honenberg (WHH) relation: $H_{C2}(0)= -0.69\, T_C(dH_{C2}/dT)$, [19] where the $T_C$ values are listed above. The linear slopes (near $T_C$) are -0.48(1) and -0.17(1) T/K and the estimated $H_{C2}^{a}(0)$ and $H_{C2}^{c}(0)$ values are 11.6(1) and 4.1(1) T for *a*- and *c*- crystals respectively. This yields the anisotropy of $H_{C2}^{a}(0)/H_{C2}^{c}(0) = 2.8(2)$. It should be noted that the WHH equation is just a rough estimation for $H_{C2}(0)$. An accurate value can only be achieved by applying high enough magnetic fields. The value obtained for $H_{C2}^{c}(0)$ is in fair agreement with $H_{C2}^{c}(0) = 5.2$ T for LSCO in the same geometry. [3] On the other hand, our $H_{C2}^{a}(0)$ is much smaller than 28 T [20] and 75 T reported in Ref. 3.

In many publications $H_{C2}(T)$ was deduced from resistivity and/or from ac susceptibility measurements. In most high-$T_C$ SC (HTSC) thin crystals, these studies provide accurate $H_{C2}(T)$ values only when H is *perpendicular* to the **wide** sample surface. [7] In this geometry the bulk nucleation of SC starts indeed at H< $H_{C2}$. On the other hand, for H *parallel* to the **wide** surface, $H_{C2}(T)$ can be deduced from bulk measurements, such as dc M(H) and/or M(T), as well as from specific heat studies. Since the method to determine $H_{C2}(0)$ in Ref. 3 is not reported, we may suspect that the high $H_{C2}^{a}(0)$ in Ref. 3, is the third critical field, $H_{C3}^{a}(0)$. Due to the limited number of publications on $H_{C2}(T)$ for La$_{1.85}$Sr$_{0.15}$CuO$_4$, any further discussions are needless. We encourage a revised estimation of the already reported $H_{C2}(T)$ values in other HTSC materials.

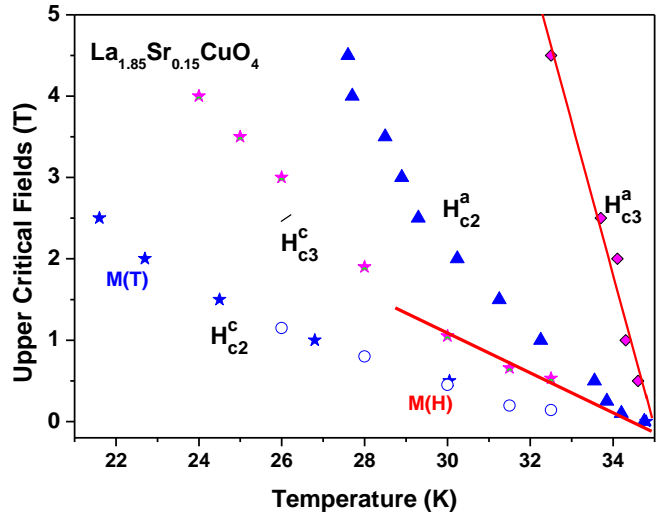

**Fig. 3**. Temperature dependence of the upper and surface critical magnetic fields: $H_{C2}^a$, $H_{C3}^a$ and $H_{C2}^c$, and $H_{C3}^c$. Note the almost linear plots near $T_C$ of all four curves. The $H_{C2}^c$ curve was obtained by M(T) (filled) and M(H) (open) methods (see text).

### (ii) Determination of $H_{C3}$(T)

The $H_{C3}$(T) values were obtained from ac susceptibility studies measured at various frequencies up to ω/2π=1465 Hz. A comparison between the ascending dc M(H) plot and the normalized χ' (taken at 1465 Hz) for H∥c of the c-crystal both measured at 26 K, are shown in Fig. 4. As stated above, the dc curve yields the $H_{C2}$ value. It is readily observed that χ', although measured under the same conditions, demonstrates clearly the presence of SC at H> $H_C$ and $H_{C3}$ is determined by its disappearance. Fig. 5 depicts the real and imaginary ac plots of the c-crystal (for H∥c) at three typical temperatures, from which the $H_{C3}^c$(T) curve in Fig. 3 was constructed. The same approach was applied to the a-crystal.

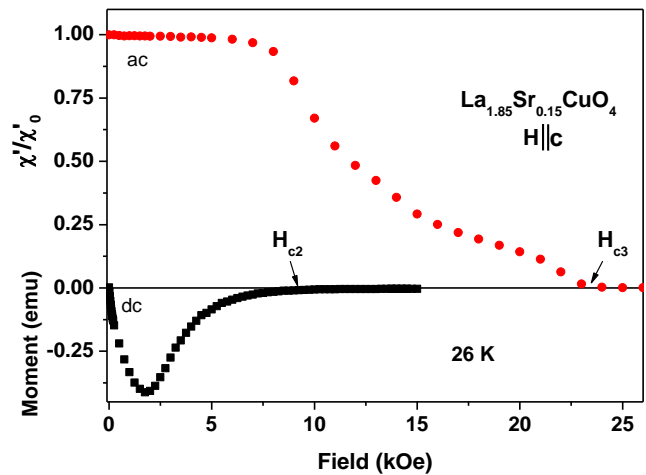

**Fig. 4**. A comparison between the dc magnetization and the normalized ac χ' susceptibility measured at 26 K under the same experimental conditions for H∥c of the c-crystal, from which $H_{C2}$ and $H_{C3}$ are deduced.

Here again, near $T_C$, both $H_{C3}^a$(T) and $H_{C3}^c$(T) plots are almost linear, with the slopes of $dH_{C3}/dT$ = -1.95(1) and -0.31(1) T/K respectively. This indicates that $\gamma^a = H_{C3}^a / H_{C2}^a$ = 4.0(2) and $\gamma^c = H_{C3}^c / H_{C2}^c$ =1.8(2). Within the uncertainty values, this $\gamma^c$ fits well the predicted γ=1.69 as discussed above. [4] On the other hand, the unexpectedly high $\gamma^a$ value is very similar to that obtained for the same orientation in the layered $K_{0.73}Fe_{1.68}Se_2$ crystal. [8]

To the best of our knowledge, no theory exists for such high $\gamma^a$ in LSCO. This high $\gamma^a$ cannot be explained by the multiband structure and standard boundary conditions of the order parameter, because for this model, $\gamma^a$ does not exceed 1.75 as demonstrated in our previous publication [21]. We may speculate that *non-standard* boundary conditions of the order parameter are responsible for this phenomenon. [7]

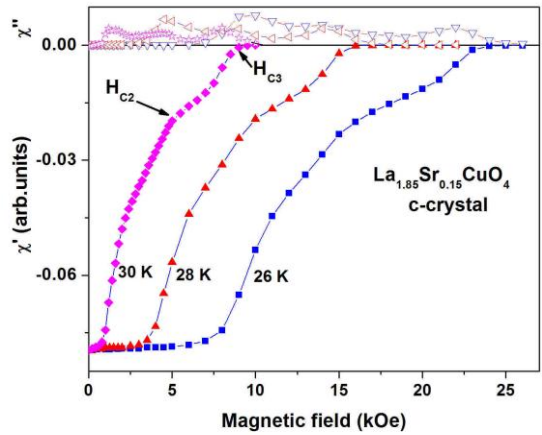

**Fig. 5.** Isothermal field dependence of the real and imaginary ac susceptibility branches of the *c*-crystal measured at 26, 28, 30 K, from which $H_{C3}^c$(T) was deduced.

### (iii) The Paramagnetic Meissner effect (PME)

*The c-crystal*. The dc ZFC and FC magnetization curves of the *c*-crystal were measured parallel (at 2.3 Oe) and perpendicular (at 1.4 Oe) to the longer c- axis. Fig. 6 (a) shows the normalized ZFC and FC M/H branches obtained in both orientations. As expected, the two ZFC branches are diamagnetic. Due to the different demagnetization factors (not calculated) the shielding fractions in the two directions are slightly different. On the other hand, in the FC process, for H||*c* the expected diamagnetic Meissner state below $T_C$ is obtained, whereas for H||*ab* the signals are positive. This, anisotropy indicates that a PME is obtained for the H||*ab* direction **only**. It is worth noting, that increasing the cooling rate by an order of magnitude did not affect the PME signals. The shielding fraction, deduced from the ZFC branch (for H||*ab*) accounts for 96%, indicating a rather perfect bulk superconductor. As a controlled experiment, we have crushed one of crystals into powder and measured the ZFC and FC curves at low H. As expected, all FC branches are negative as depicted for the H||*c*.

*The a-crystal*. The same PME is observed in the a-crystal and for H||*ab* (the long dimension). (Fig. 6(b)). In order to check the reproducibility of the anisotropy in the PME, we measured the a-crystal along its two short dimensions. In contrast to the *c*- crystal, for H applied **perpendicular** to the long dimension, two possibilities are present. (i) H may be applied along the *crystallographic c*-axis which is shortest dimension (1.8 mm). (ii) Alternatively H may be

applied parallel to the ab basal planes (the middle dimension (2.2 mm)). Fig. 7(a) shows that in the first case, for H||c all FC curves obtained measured up to H=53 Oe are negative, exhibiting the regular typical Meissner effect. On the other hand, in the second case, for H||ab, the PME phenomenon is readily observed (Fig. 7(b)). This definitely proves that in both a and c crystals, the PME is observed only when H is applied along the basal planes.

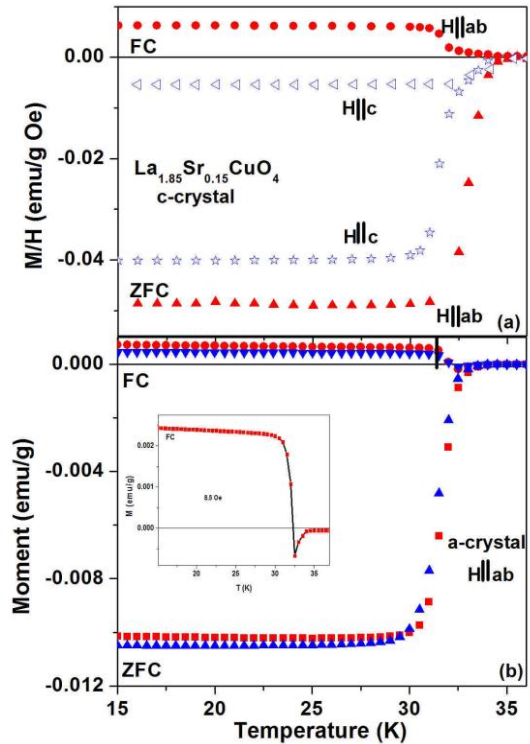

Fig. 6. (a) ZFC and FC susceptibility plots of the c-crystal for H applied parallel and perpendicular to the c- axis measured at 2.3 and 1.4 Oe respectively. (b) ZFC and FC magnetization curves of the a-crystal, measured at 3.5 (blue) and 5.5 Oe (red) for H parallel to the long axis. The inset shows the FC plot measured at 8.5 Oe

Remarkably, the observed PME in various SC systems which appears only at very low H values (less than 1 Oe), is exhibited here up to H of ~10–15 Oe. For the a-crystal, the positive M/H (measured at 25 K along the long dimension), decreases with H as $M/H = C*H^{-\alpha}$, where the constant C=0.0021 and $\alpha = 1.0 \pm 0.05$, indicating that the M/H is inversely proportional to H (Fig. 7(b) inset).

Moreover, due to the similarity in the short and middle dimensions, we may assume the in both directions the demagnetization factors are quite similar; thus we may compare between the normalized positive and negative signals presented in Fig.7(b). As expected, the two ZFC curves obtained are quite similar in both directions. On the other hand in the FC branches, the positive PME signal is twice as much as in the diamagnetic one.

In conclusion, for LSCO crystal, (i) the PME is observed only for H parallel to the Cu-O layers, (ii) the positive signal in larger than the Meissner state. This observation differs strictly from that reported for YBCO crystal, where the PME signal is small and was observed only for H normal to the Cu-O layers. [9]

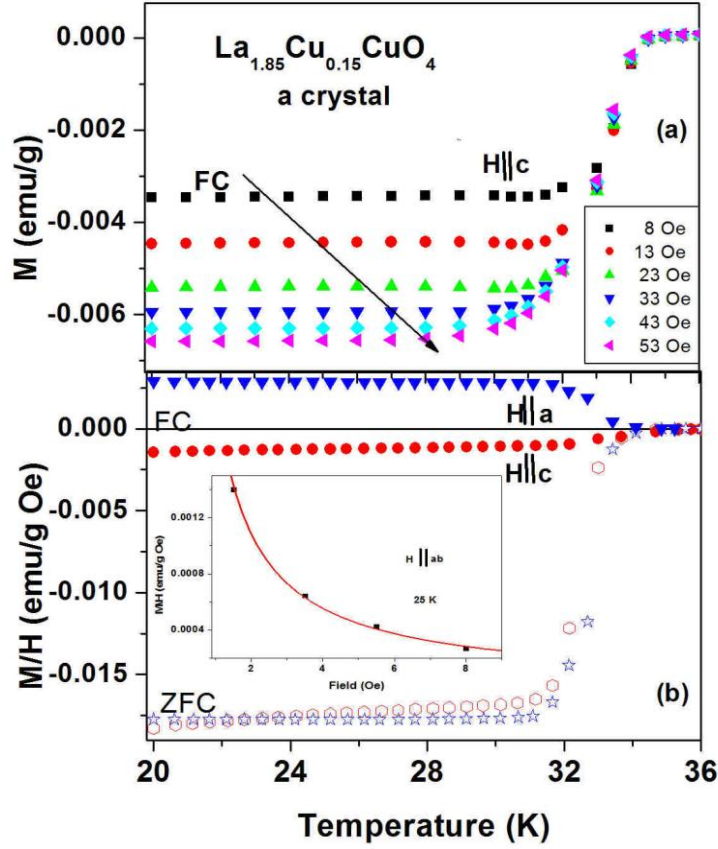

Fig.7. (a) FC plots of the *a*-crystal for H||*c*. (b). The normalized ZFC and FC plots of the a-crystal measured perpendicular to the long axis, for H parallel and perpendicular to the ab planes measured at 4 Oe (red) and 2.5 Oe (blue) respectively. The inset shows the field dependence of the PME effect for H||*a*.

### Discussion

Two important physical parameters of the optimal doped $La_{1.85}Sr_{0.15}CuO_4$ single crystal are addressed here: (i) We show the phase diagram which includes the upper critical fields $H_{C2}(T)$ and $H_{C3}(T)$ for both H||*c* and H||*ab* orientations; (ii) the PME phenomenon which is observed only when H is parallel to the Cu-O basal planes. To our best knowledge, these observations are reported for the first time.

(i) In contrast to YBCO, only a limited number of publications have reported on the upper critical field of LSCO. [20-23] Due to the lack of large single crystals most of the published results report mainly on the doping dependence of $H_{C2}(T)$ along the *c*- axis. It is well accepted that $H_{C2}(T)$ is anisotropic and that $H_{C2}^a > H_{C2}^c$, however the anisotropy ratio 2.8(2) presented here, is smaller than 5.4 reported in Ref. 3. By using the ac susceptibility technique, we studied the surface critical magnetic fields $H_{C3}$(T) and deduced the ratio $\gamma = H_{C3}/H_{C2}$ in both H||c and H||ab directions. $\gamma^c = 1.8(2)$, fits well the theoretical 1.69 value. On the other hand, a much higher $\gamma^a = 4.0(2)$ is obtained.

(ii) The theoretical models of the PME were discussed earlier in several publications. [14–16]. There is no consensus as to what are the essential factors which cause the PME. It is proven here, that for LSCO, this anisotropic phenomenon is observed only when H is parallel to the Cu-O planes regardless of whether it is measured parallel to the long (a-crystal) or the short (c-crystal) dimension of the crystal. Moreover, Fig. 7 shows that, perpendicular to the long dimension of the *a*-crystal, the PME is observable for H||ab only but not and H||c although the demagnetization factors for both directions are almost the same. That excludes the assumption that demagnetization effects enhance the PME [11], or that the PME is caused by impurities which act as effective pinning centres for the vortices [16]. Worth mentioning, in contrast to YBCO single crystals,[9] in LSCO, the PME signal is larger than the negative Meissner branch (Fig. 6).

The theoretical model for the PME is beyond the scope of the present study. However, we may suggest that the two phenomena observed: the higher $\gamma^a$ ratio and the PME along the Cu-O planes are practically related to each other. For the layered LSCO material, $H_{C2}^c = \Phi_0/2\pi\xi_{ab}^2$ and $H_{C2}^a = \Phi_0/2\pi\xi_{ab}\xi_c$ where $\Phi_0$ is the magnetic flux quantum and $\xi_{ab}$ and $\xi_c$ are the coherence lengths parallel to the Cu-O planes and the *c*- axis respectively. Thus, the fact that $H_{C2}^a > H_{C2}^c$ is caused by the anisotropy of the coherence length ($\xi_{ab} > \xi_c$). We may speculate that for H applied parallel to the Cu-O basal planes, the two coherence lengths cause two types of induced currents, which are exhibited by the PME and by higher $\gamma^a$ value. This is in line with the model which connects the PME to the surface superconductivity, when $T_C$ of the surface is different from that of the bulk $T_C$ (see Fig. 1). [6,24] However, the current state of experiments does not allow us to suggest any consistent explanation for these anisotropies. It is possible that there is some interplay between them, but no existing theoretical models that would be able to explain these magnetic phenomena have been proposed. We encourage the development of a new model which will take these observations into account.

*In summary,* we have demonstrated the existence of surface superconductivity in rectangular needle-like shapes of $La_{1.85}Sr_{0.15}CuO_4$ single crystals. DC magnetization measurements yield an anisotropic ratio of 2.8(2) of the bulk upper critical fields $H_{C2}(0)$ when measured along the basal planes and the *c*-axis. From the ac susceptibility study we have deduced the magnetic critical field of the surface $H_{C3}$(T) where ($H_{C3} > H_{C2}$). In the *c*-axis direction $H_{C3}^c/H_{C2}^c = 1.8(2)$, a value which is in fair agreement with the theoretical 1.69 ratio. On the other hand, an unexpected high ratio $H_{C3}^a/H_{C2}^a = 4.0(2)$ is obtained in the Cu-O planes. In addition, we observed an anisotropic PME. Positive FC branches were obtained only for H parallel to the Cu-O basal planes, regardless the dimension of the crystal. It is speculated that the two anisotropies are connected to each other.

**Acknowledgments**: The research in Jerusalem and in the Technion is supported in part by the Israel Science Foundation ISF (389/09), ISF Bikura grant (459/09), the German-Israel DIP program and by the Klachky Foundation for Superconductivity. M.I.T. thanks V.M. Genkin for valuable discussions.

*israela@vms.huji.ac.il